\documentclass[12pt]{iopart}
\usepackage{color}
\usepackage{graphicx}
\usepackage{amsmath, amsthm, amssymb}
\usepackage{enumerate}
\usepackage[normalem]{ulem}
\definecolor{Gray}{gray}{0.9}
\usepackage{citesort}
\usepackage{tabularx}

\newcommand{\be}{\begin{equation}}
\newcommand{\ee}{\end{equation}}
\newcommand{\bea}{\begin{eqnarray}}
\newcommand{\eea}{\end{eqnarray}}

\begin{document}
\title{The High Density Phase of the $k$-NN Hard Core Lattice Gas Model}
\author{Trisha Nath$^1$
 and R. Rajesh$^2$}
 \ead{$^1$trishan@imsc.res.in,$^2$rrajesh@imsc.res.in}
 \address{The Institute of Mathematical Sciences, C.I.T. Campus,
 Taramani, Chennai 600113, India}

\date{\today}

\begin{abstract}
The $k$-NN hard core lattice gas model on a square lattice, in which the 
first $k$ next nearest neighbor sites of a particle are excluded from being occupied 
by another particle, is the lattice version of the hard disc model in two dimensional continuum.
It has been conjectured that the lattice model, like its continuum counterpart, will show multiple
entropy-driven transitions with increasing density if the high density phase has columnar or striped
order.
Here, we determine the nature of the phase at full packing for $k$ up to $820302$. We show that
there are only eighteen values of $k$, all less than $k=4134$, that show columnar order, while
the others show solid-like sublattice order.
\end{abstract}
\maketitle

\section{\label{I}Introduction}

Models of particles interacting  through  only hard core interactions are the simplest
theoretical models to show phase transitions. All order-disorder transitions seen
in such models are geometrical and entropy driven. A well-known example is the 
system of hard 
spheres~\cite{alderdisc,aldersphere}.  In three dimensions,  it
undergoes a discontinuous transition from a liquid phase to a solid phase 
with increasing packing fraction. 
In two dimensions, the freezing occurs in two steps:
first from a liquid phase to a hexatic phase with quasi long range orientational order and
second from the hexatic phase to a solid phase with orientational order and quasi long range positional
order~\cite{kosterlitz,nelson,yong,krauth2011,wierschem2011}.

The lattice version of the hard-sphere model, the $k$-NN model, has also been extensively
studied since being introduced  in the 1950s~\cite{domb,burley1,burley2} (also 
see Refs.~\cite{fernandes,trisha} and references within for other applications).
In this model, the first $k$ next nearest neighbors of a particle are excluded from being
occupied by another 
particle. 
On the triangular lattice, the 1-NN model reduces to the hard hexagon
model, one of the few exactly solvable hard core lattice gas models~\cite{baxter}. In this paper, we focus on the continuum limit of the $k$-NN model on the square lattice. This corresponds
to the limiting case  of large $k$, when it might be 
expected that the  model converges to the hard sphere system in two
dimensions.

We briefly summarize what is known about the different phases and phase transitions 
observed in the $k$-NN model on a square lattice. The 1-NN model,  in which the nearest neighbor
sites of a particle are excluded, undergoes a 
continuous transition belonging to the Ising universality class 
from a low density disordered phase to a high density solid-like sublattice phase with two-fold
symmetry~\cite{gaunt_fisher,bellemans_nigam1nn,baxter_enting_tsang,baram,bellemans_nigam1nn,runnels,combs,ree2,nisbet,guo,pears,chan1,jensen,landau,meirovitch,hu_mak,fernandes,liu,lafuente,racz,hu_chen}. The
high density phase of the 2-NN model, also known as the $2 \times 2$ hard square model, is columnar with translational
order present only along either rows or columns but not along both~\cite{bellemans_nigam1nn,bellemans_nigam3nn,ree,landau,nisbet1,lafuente2003phase,ramola,schmidt,amar,slotte,kinzel,rdd2015,trisha2016}. 
The ordered phase has a four-fold symmetry  and the transition is continuous and belongs to the 
Ashkin-Teller universality class~\cite{fernandes,feng,zhi2}. The 3-NN model undergoes a
discontinuous transition from the disordered to a sublattice phase with ten fold symmetry~\cite{bellemans_nigam3nn,orban1966,orban1982,bellemans_nigam1nn,nisbet2,heilmann,eisenberg2005,fiore,fernandes,nisbet1}. The 5-NN ($3 \times 3$ hard square) model undergoes a first order transition
into a columnar phase with 6 fold symmetry~\cite{fernandes}.

In a recent paper, we showed numerically that the 4-NN model surprisingly undergoes two phase transitions as density is increased: first from a low density disordered phase to an intermediate
sublattice phase and second from the sublattice phase to a high density
columnar phase~\cite{trisha}. The existence of two transitions was rationalized by
deriving a high density expansion about the ordered columnar phase.  Columnar phases have
a sliding instability in which a defect created by removing a 
single particle from the fully packed configuration splits into  fractional defects that slide independently of each other
along some direction~\cite{bellemans_nigam1nn,ramola,trisha2}.
The high density expansion for the $4$-NN model showed
that the sliding instability is present in only certain preferred sublattices. As density is decreased from full packing, it is thus plausible that the columnar phase destabilizes into these preferred sublattices resulting in a sublattice phase rather than a disordered phase. This led us to conjecture that
for a given $k$, if the model satisfies the two conditions (i) the high-density phase 
is columnar and (ii) the sliding instability is present in only a fraction of 
the sublattices, then the system will 
show multiple transitions. Implementing a Monte Carlo algorithm with cluster 
moves~\cite{kundu,kundu2,kundu3}, we were
able to numerically verify the conjecture for $k=6,7,8,9$, by showing the presence 
of a single
first order transition, and for $k=10,11$ (for which the high density phase 
is columnar), by showing the presence of multiple
phase transitions~\cite{trisha}. The results for $k\leq11$ are summarized in Table.~\ref{table:generalk}.
\begin{table}
\caption{\label{table:generalk} The nature of the  ordered phases and the transitions for $k\leq 11$.
More than one entry for  a given value of $k$  indicates that 
multiple transitions occur with increasing density  with the phases appearing in the order it is listed.

\noindent Question marks denote that the result is not known.} 
\begin{indented}
\lineup
\item[]\begin{tabular}{@{}lll}
\br
k  & Ordered phases & Nature of transition  \\
\mr
   1 &  Sublattice & Continuous (Ising)  \\
   2 &  Columnar & Continuous (Ashkin-Teller) \\
   3 &  Sublattice & Discontinuous  \\ 
   4 &  Sublattice &  Continuous (Ising)  \\ 
     & Columnar & Continuous (Ashkin-Teller) \\
   5 &  Columnar &  Discontinuous  \\ 
   6 &  Sublattice &  Discontinuous  \\ 
   7 &  Sublattice &  Discontinuous  \\ 
   8 &  Sublattice &  Discontinuous \\ 
   9 &  Sublattice &  Discontinuous \\
  10 &  Sublattice &  Continuous (Ising) \\ 
     & Columnar &  Discontinuous \\
  11 & 	? & ? \\
  & 	Columnar & ? \\
  \br
  \end{tabular}
  \end{indented}
\end{table}

The presence of multiple transitions  in the $k$-NN model, which was
hitherto unexpected, raises the intriguing possibility that  in the continuum
limit  $k\to \infty$, the system may show multiple transitions as in the hard
sphere problem in two dimensions.  Monte Carlo simulations of
systems with $k \geq 12$ is impractical as, at high densities, the large excluded volume per particle
 results in the
the system getting stuck
in long lived meta stable states, making it a poor candidate for
studying large $k$. Instead, in this paper, we assume that the conjecture stated above is 
correct (based on it working for $k=1,2, \ldots,11$), and determine those values of $k$ for
which there are multiple transitions. 

The rest of the paper is organized as follows.
In Sec.~\ref{II}, we precisely define the model and the columnar and sublattice phases. In Sec.~\ref{III}, 
we explain the procedure for determining whether the high density phase has columnar order. 
Implementing this algorithm, we show that the number of system with columnar order is finite and that for large $k$, the high density phase has sublattice order. We conclude with discussions in Sec.~\ref{IV}.

\section{Model and definitions\label{II}}

\begin{figure}
\centering
\includegraphics[width=0.5\columnwidth]{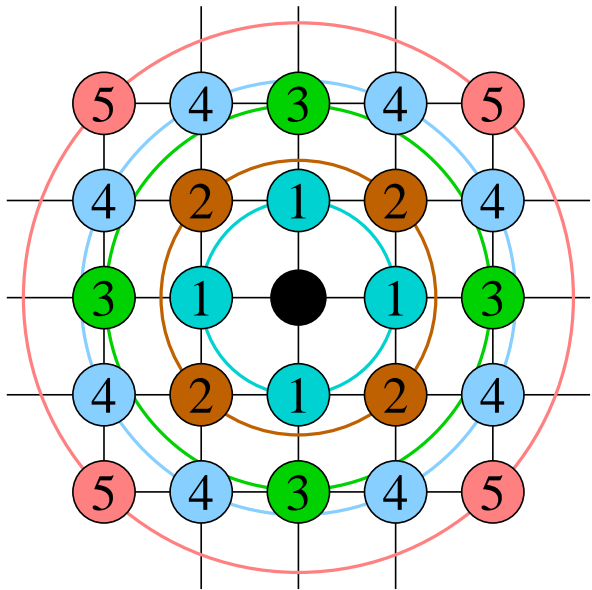}
\caption{ Consider a particle placed on the central black lattice site.  
The labels $n=1$ to $n=5$ denote the sites 
that are the $n^{th}$ next nearest neighbors of the particle.  In the $k$-NN hard core lattice
gas model all the sites with labels less than or 
equal to $k$ are excluded from being occupied by another particle. The 
excluded regions are also shown by concentric circles such that
all sites within the circle of radius $R$ are excluded for a given $k$.
}
\label{fig:knn}
\end{figure}

Consider a square lattice with periodic boundary conditions. A site may be empty 
or occupied by utmost one particle. A particle in a $k$-NN model excludes the first $k$ next nearest neighbors 
from being occupied by another particle.  Figure~\ref{fig:knn} shows the 
excluded sites  for $k=1,\ldots5$, where the label $n$ refers to the $n$-th next nearest neighbor. We refer to this model
as the $k$-NN model.  As an example, only the nearest neighbors are excluded in the 1-NN model.  
With increasing $k$, the successive excluded regions correspond to lattice sites within  
circles of radius $R$, where $R^2$ are
norms of the Gaussian integers.

Our conjecture states that for multiple phase transitions to be seen,
the high-density phase should be columnar in nature and 
the sliding instability should be present in only a fraction of 
the sublattices. If the excluded volume is a perfect square, then
the high density phase will be columnar, but the sliding instability
will be present in all sublattices. However, the excluded volume is a
hard square only when $k=2$ and $k=5$. Thus, the conjecture reduces
to determining for a given $k >5$ whether the high density phase has
columnar order. This may be determined by examining the fully-packed
phase.

\begin{figure}
\centering
\includegraphics[width=0.8\columnwidth]{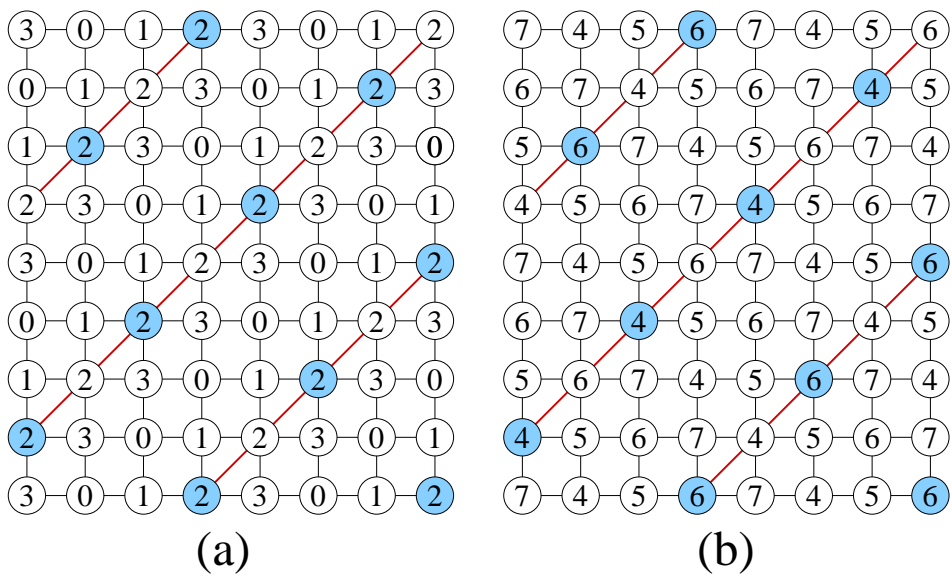}
\caption{  The sublattices for the  4-NN model. The lattice sites are labeled according to
diagonals oriented in the (a) $\pi/4$ direction or (b) $3 \pi/4$ direction. The filled blue sites
correspond to a typical configuration at maximal density. In the example shown the particles
are in sublattice $2$ [see (a)] or equivalently sublattices $4$ and $6$ [see (b)].
}
\label{fig:sublattice}
\end{figure}

We now give an example of a columnar phase at full packing to identify
a criterion to determine whether the fully packed phase has columnar order.
Consider the 4-NN model whose high density phase has columnar order~\cite{trisha}. A
typical configuration at full packing is shown in Fig.~\ref{fig:sublattice}. We divide the lattice into four sublattices depending on the diagonal that it belongs to. However, diagonals
may be oriented in the $\pi/4$ [see Fig.~\ref{fig:sublattice}(a)] or  in the 
$3 \pi/4$ [see Fig.~\ref{fig:sublattice}(b)]
directions and, thus, two such labellings are possible. In the configuration shown, all particles
are in sublattice $2$ when sites are labeled from $0$ to $3$ [see Fig.~\ref{fig:sublattice}(a)],
but in sublattices $4$ and $6$ when sites are labeled from $4$ to $7$ [see Fig.~\ref{fig:sublattice}(b)].
Clearly, there are $8$ such phases possible. The key feature is that when boundary conditions
are periodic, then the particles may slide freely along diagonals (red lines in Fig.~\ref{fig:sublattice}), independent of other diagonals.
Thus, the degeneracy of the fully packed phase increases exponentially with system size.

\begin{figure}
\centering
\includegraphics[width=0.5\columnwidth]{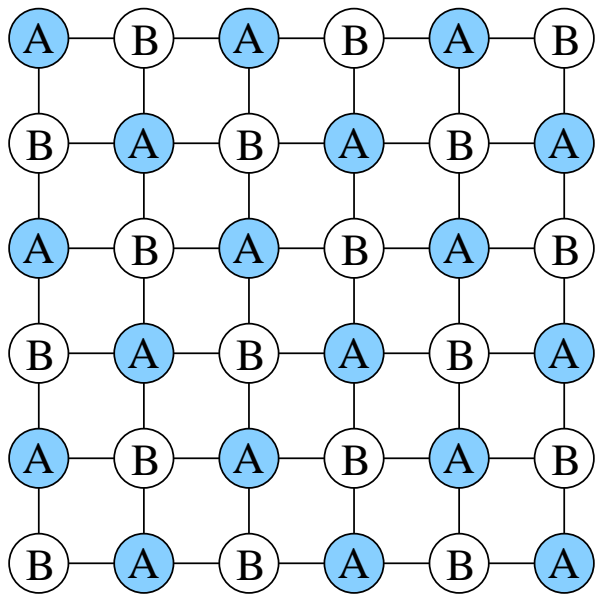}
\caption{ The two sublattices  for the 1-NN model are shown by $A$ and $B$.  
A fully packed configuration (shown by blue filled sites) consists of all particles
being in sublattice $A$ (as shown in figure) or in sublattice $B$.} 
\label{fig:1nn}
\end{figure}

In contrast, when the system has sublattice order,  the number of fully packed configurations
are finite. An example is the 1-NN model. In this case, the lattice may be divided into
two sublattices such that neighbors of a site of  one sublattice  belong to the other sublattice (see
Fig.~\ref{fig:1nn}). In the limit of full packing, all the particles are either in sublattice $A$ or in 
sublattice $B$, and only two configurations are possible irrespective of system size.

Thus, we fix the the criterion for columnar order at full packing to be that the particles
should be slidable along some direction (not necessarily $\pi/4$ or $3 \pi/4$ as in 4-NN)
independent of the positions
of the other particles resulting in a highly degenerate fully packed state.
In Sec.~\ref{III},  we describe the algorithm for checking slidability in the fully packed
configuration.

\section{Results\label{III}}

\begin{figure}
\centering
\includegraphics[width=0.5\columnwidth]{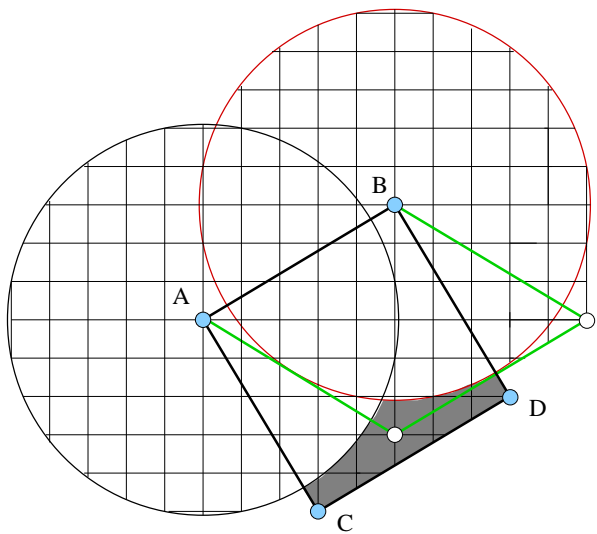}
\caption{ A schematic diagram to explain the procedure for constructing the unit
cell at full packing. The unit cell is a parallelogram ABCD. The 
blue (filled) circles are particles. All 
lattice points inside the black and red  circles are excluded by $A$ and $B$ respectively. 
If $C$ is placed on the lattice site denoted by white (open) circle, then the parallelogram
has smaller area than the square $ABCD$.}  
\label{fig:structure}
\end{figure}

To find the possible configurations for the k-NN model
at full packing, we proceed as follows. 
The unit cell that repeats to give the fully packed configuration is a
parallelogram. 
Let the particle at the top left corner of the 
parallelogram be denoted by $A$ (see Fig.~\ref{fig:structure}). 
We choose the origin
of the coordinate system to be at $A$. The excluded volume of $A$ is a 
circle (shown in black in Fig.~\ref{fig:structure}) whose radius
is dependent on $k$. Let the particle at the top right corner of
the parallelogram be denoted by $B$ with coordinates $(x,y)$.
We restrict the choice
of $B$ to be in the first octant ($x\geq y$), as a choice in the second
octant may be mapped onto the first octant by rotation.
For every $y\geq 0$, $x$ is the minimum value such that $x^2+y^2>R^2$,
where $R$ is the radius of the excluded volume.
The excluded volume of $B$ is shown by a red circle in 
Fig.~\ref{fig:structure}.

For a fixed $A$ and $B$, the orientation and length of the segment
$CD$ is
fixed as the unit cell is a parallelogram. The position of $C$ is determined by the constraint 
that the area of the parallelogram is the minimum and is determined
as follows. A convenient initial choice of $C$ that does not violate the hard-core constraint
is obtained by rotating the
point $B$ counterclockwise about $A$ by $\pi/2$ such that the unit cell is a square (see
Fig.~\ref{fig:structure}). 
A choice of $C$ where $\angle{BAC}>\pi/2$
may be mapped onto a unit cell where $\angle{BAC}<\pi/2$ by translating the segment CD
by its length along its length. It is now straightforward
to see that any choice of $C$ that results in a smaller area for the
unit cell as compared to the initial square must lie within the initial square and outside the 
excluded volume of $A$ and $B$ (for example, the lattice point shown by an 
empty circle in Fig.~\ref{fig:structure}). We fix $C$ by minimizing the area of
the parallelogram over all such points.

The minimization of area is repeated over all  possible choices of $B$ to determine 
the unit cell and the maximal packing density. If the segment $CD$  now passes through 
any lattice point that is outside the
excluded volume of $A$ and $B$, then $C$ could be moved  to that lattice point to give a parallelogram
of same area.  Likewise, if $AC$  passes through any lattice point that is outside the
excluded volume of $B$ and $D$, then  $A$ could be moved  to that lattice point to give a parallelogram
of same area. If such multiple choices are possible, then we identify the high-density phase
to have columnar order.

\begin{figure}
\centering
\includegraphics[width=0.8\columnwidth]{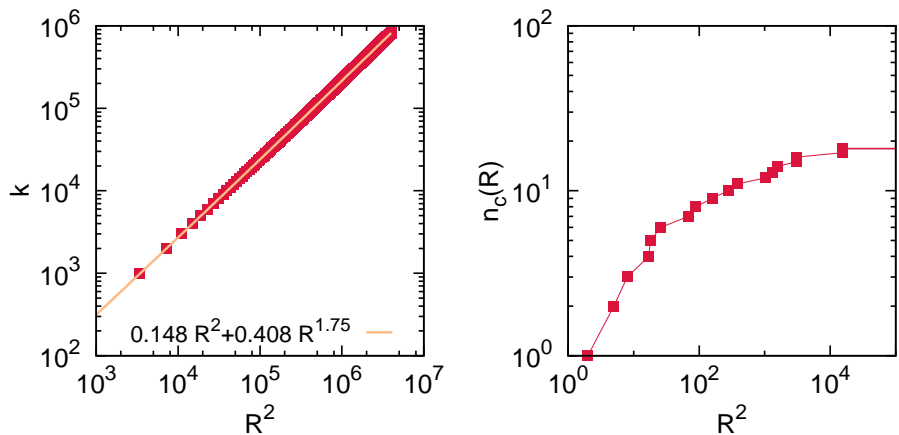}
\caption{The variation with $R^2$  of (a) $k$ and (b) $n_c(R)$.  }
\label{fig:densityrho}
\end{figure}

Using the above procedure, we identify the fully packed configuration for $k$ up to $820302$
corresponding to $R^2=3999997$, where $R$ is the radius of the smallest circle 
around a site that
encloses all its excluded lattice sites. The number of distinct values of $k$ 
increases linearly with $R^2$ [see Fig.~\ref{fig:densityrho}(a)]. Asymptotically we find
$k\approx0.148 R^2$, for $R\gg 1$. The values of $k$ for which the high density phase has columnar order are listed in Table~\ref{table:columnar}. We find that up to the values of $R^2$ that we have checked, the largest $k$ for which there is columnar order at full packing is $k=4183$ or $R^2=15482$. Among these, the stability of columnar phase for density close to full packing has been numerically established only for $k=2,4,5,10,11$~\cite{trisha,ramola,fernandes}. Let $n_c(R)$ denote the number of systems
 whose exclusion is less than or equal to $R$ and has columnar order at full packing. 
 $n_c$ increases irregularly with $R$ and saturates at $n_c=18$ 
 [see Fig.~\ref{fig:densityrho}(b)].

\begin{table}[hb]
\caption{\label{table:columnar} The values of $k$ for which the high density phase
of the $k$-NN model has columnar order. $R$ is radius of the smallest circle that
encloses the excluded lattice sites [see Fig.~\ref{fig:knn}]}.
\centering
\lineup
\begin{tabular*}{0.5\textwidth}{@{\extracolsep{\fill} }l l }
\br
k & $R^2$ \\
\mr
   2 &  2  \\
   4 &  5  \\
   5 &  8 \\ 
   10 &  17  \\ 
   11 &  18 \\ 
   14 &  26 \\ 
   31 &  68  \\ 
   39 &  89  \\ 
   64 &  157  \\ 
  105 &  277  \\ 
  141 &  389\\
  342 &  1040  \\
427 & 1322 \\
493 & 1557 \\
906 & 3029\\
907 & 3033\\
4132 & 15481\\
4133 & 15482\\
\br
  \end{tabular*}
\end{table}

\begin{figure}
\centering
\includegraphics[width=0.9\columnwidth]{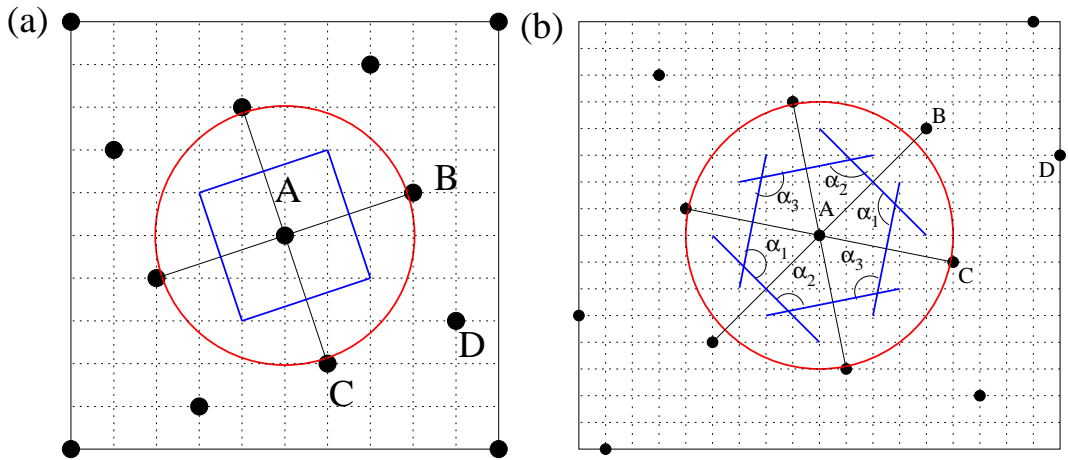}
\caption{ Wigner-Seitz cells for $6$-NN and $13$-NN models. Black dots are particles. The surrounding red circles are the exclusion regions of 
the central particle A. Blue lines are the perpendicular bisectors on the lines joining neighboring 
particles to $A$. Wigner-Seitz cell is a (a) square for $6$-NN, (b) hexagon for $13$-NN. }
\label{fig:example}
\end{figure}

For large  $k$, the fully packed configuration has sublattice or crystalline order. 
For each of this $k$ we construct the Wigner-Seitz primitive cell obtained by constructing the convex envelope of the perpendicular bisectors of the lines joining a fixed particle to all other particles in the fully packed configuration.  
 If the unit cell $ABCD$ is such that  $\angle BAC$ equals $\pi/2$, then the 
Wigner-Seitz cell and the unit cell $ABCD$, are both squares.  
We find that the Wigner-Seitz cell is a square only for $k\leq6$ and for $k=11$. 
The Wigner-Seitz cell for $6$-NN is shown in Fig.~\ref{fig:example}(a).

For other values of $k$, we observe that $\angle BAC$ of the unit cell is always smaller than $\pi/2$. 
Also the length of the two sides, $AB$ and $AC$, are unequal.
For these $k$, the Wigner-Seitz cell is  an irregular hexagon where the opposite sides are parallel and of 
same length. Thus, pairs of opposite angles of the hexagon  are equal to each other. As an example,
the Wigner-Seitz cell for $13$-NN is shown in Fig.~\ref{fig:example}(b).

\begin{figure}
\centering
\includegraphics[width=20cm,height=20cm,keepaspectratio]{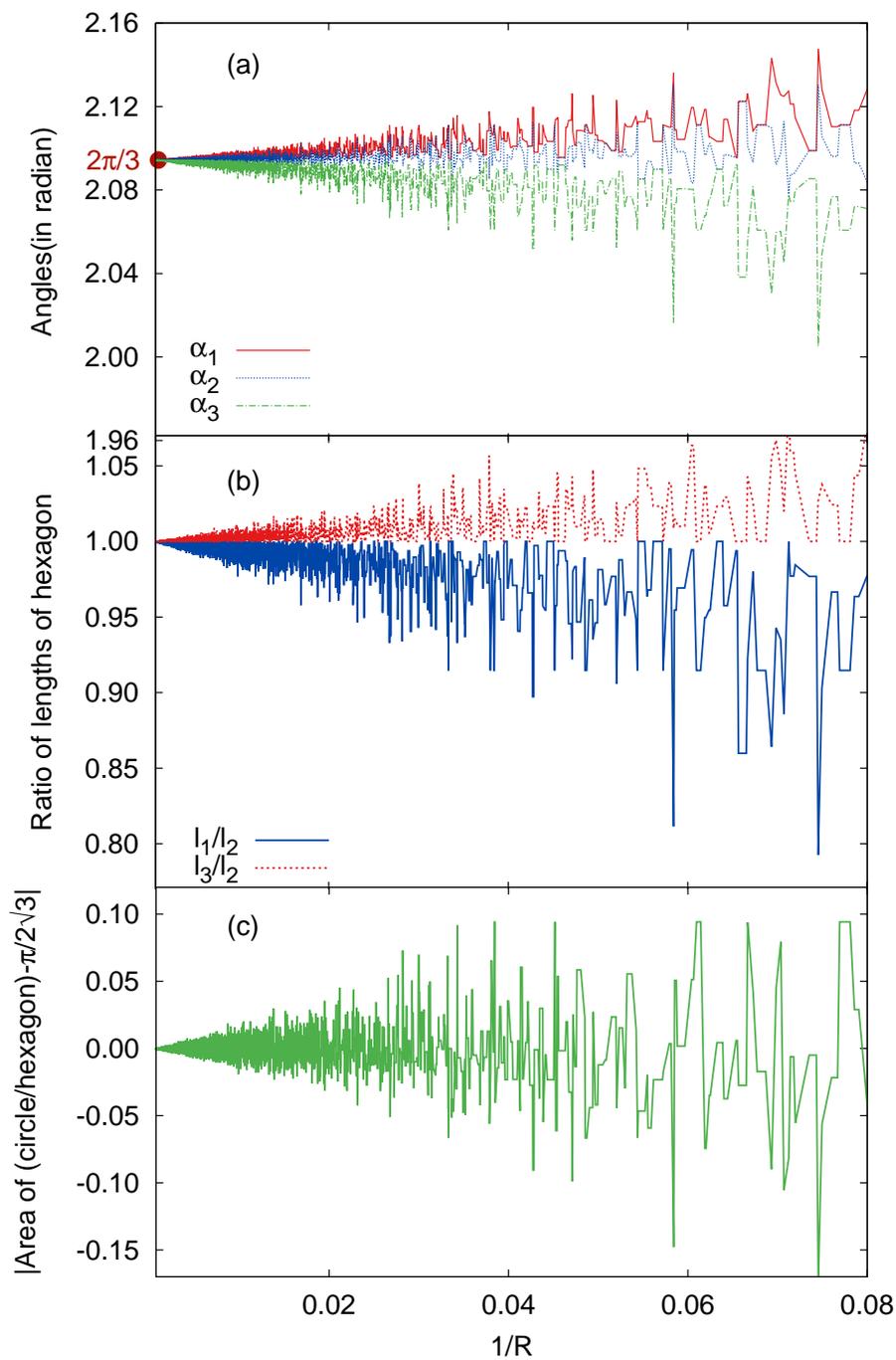}
\caption{The dependence on exclusion radius $R$ of 
(a) angles $\alpha_i$, 
(b) ratios of lengths of different sides, and
(c) ratio of area of the circumscribed circle to  area of the
hexagonal Wigner-Seitz cell.
}
\label{fig:angle}
\end{figure}

We now characterize the shape of the Wigner-Seitz cell. Let the angles of the hexagon
be denoted by $\alpha_1$,  $\alpha_2$ and $\alpha_3$, such that  
$\alpha_1 \leq \alpha_2 \leq \alpha_3$. As $R\to \infty$,  the three angles approach  $2\pi/3$ linearly 
with decreasing $1/R$ (see  Fig.~\ref{fig:angle}(a)). 

Likewise, let $l_1$, $l_2$, and $l_3$ denote the lengths of the hexagon 
such that $l_1\leq l_2\leq l_3$.
The ratios $l_1/ l_2$ and $l_3/ l_2$ approach $1$ linearly with decreasing $1/R$ (see  Fig.~\ref{fig:angle}(b)). We thus conclude that the Wigner-Seitz cell converges to a regular hexagon with increasing $k$.

We also check that in the limit of large $R$, the packing approaches that of discs in two dimensions.
For discs, the largest packing density is achieved when the packing is hexagonal close packing
with packing density $\eta_h=\pi/(2\sqrt{3})$. 
In the limit of large $R$,  the ratio of the area of the largest circle circumscribed in the hexagonal Wigner-Seitz cell  to the area of the hexagonal cell approaches $\eta_h$ (see Fig.~\ref{fig:angle}(c)).

\section{Conclusions\label{IV}}

 In this paper we obtained the nature of the phase at maximal density for the $k$-NN model hard core lattice gas models in which the 
first $k$ next nearest neighbor sites of a particle are excluded from being occupied 
by other particles by finding out numerically the configuration that maximizes the density at full packing. We
find that for up to  $k=820302$, there are only 18 values of $k$ for which the high density phase has
columnar order. The remaining ones have solid-like sublattice order. For these systems, we showed that
the Wigner-Seitz primitive cell approaches a regular hexagon for large $k$, and the packing
tends to a hexagonal packing.

The largest value of $k$ for which we found  columnar order is $k=4133$. Since this is 200 times smaller than the maximum value of $k$ that we have tested, it appears reasonable to conclude the number
of systems with columnar order is finite.

If the conjecture that multiple transitions exist only if the high density phase is columnar~\cite{trisha}
is true, then we conclude that for large $k$, the system shows one first order  transition to a phase with
sublattice order.  In this case, the analogy to the hard disc problem in the two dimensional continuum,
for which there are two transitions, breaks down. It would be interesting to test the conjecture more 
rigorously. One possibility is that even for systems with sublattice order, there could be two transitions.
To show this, one needs to construct the high density expansion about the sublattice phase for large
$k$ and check whether certain defects are preferred over others.

\section*{References}

 \providecommand{\newblock}{}

\end{document}